\documentclass{article}%
\usepackage{amssymb}
\usepackage{amsmath}
\usepackage{amsfonts}
\usepackage{graphicx}%
\begin{document}

\title{Nonlinear generalized master equations and accounting for initial correlations}
\author{Victor F. Los$^{\ast}$\\Institute of Magnetism, Nat. Acad. of Sci. of Ukraine, \\36-b Vernadsky Blvd., Kiev 03142, Ukraine}
\maketitle

\begin{abstract}
By using a time-dependent operator converting a distribution function
(statistical operator) of a total system under consideration into the relevant
form, new exact nonlinear generalized master equations (GMEs) are derived. The
inhomogeneous nonlinear GME is a generalization of the linear Nakajima-Zwanzig
GME and is suitable for obtaining both the linear and nonlinear equations
describing the evolution of the (sub)system of interest. Actually, this
equation provides an alternative to the BBGKY chain. To include initial
correlations into consideration, this inhomogeneous nonlinear GME has been
converted into the homogenous form by the method suggested earlier in
\cite{Los (2001), Los (2005)}. Obtained in such a way exact homogeneous
nonlinear GME describes all stages of the (sub)system of interest evolution
and influence of initial correlations at all stages of the evolution. The
initial correlations are treated on the equal footing with collisions and are
included in the modified memory kernel acting on the relevant part of a
distribution function (statistical operator). No approximation like the
Bogoliubov principle of weakening of initial correlations or RPA has been
used. In contrast to homogeneous linear GMEs obtained in \cite{Los (2001), Los
(2005)}, the homogeneous nonlinear GME is convenient for getting both a linear
and nonlinear evolution equations. The obtained nonlinear GMEs have been
tested on the space inhomogeneous dilute gas of classical particles.
Particularly, a new homogeneous nonlinear equation describing an evolution of
a one-particle distribution function at all times and retaining initial
correlations has been obtained in the linear in the gas density approximation.
This equation is closed in the sense that all two-particle correlations
(collisions), including initial ones, which contribute to dissipative and
nondissipative characteristics of the nonideal gas are accounted for in the
memory kernel. Connection of this equation at the kinetic stage of the
evolution to the Vlasov-Landau and Boltzmann equations is discussed.



$^{\ast}$Tel/Fax: +38-044-253-8821.

\textit{E-mail address:} vlos@imag.kiev.ua

\end{abstract}

\section{Introduction}

Derivation of equations allowing for describing the evolution of the
measurable (statistical expectation) values, which characterize a
non-equilibrium state, remains one of the principal tasks of statistical
physics. To get kinetic (or other irreversible) equations from underlying
microscopic (reversible) classical or quantum dynamical equations, several
approaches are usually used, which generally start with the
Liouville-von-Neumann equation for a distribution function (statistical
operator) of the many-body system under consideration.

One of them leads to the chain of coupled first-order differential equations
for many-particle distribution functions known as the BBGKY hierarchy
\cite{Bogoliubov (1946)}. In order to obtain from the BBGKY hierarchy a closed
kinetic equation for a reduced (e.g. a one-particle) distribution function,
one needs to decouple the chain in some approximation using a small parameter
such as an interaction between particles or a density of the system $n$.
Actually, it may be done by disregarding \textbf{all} initial (existing at the
initial time moment $t_{0}$) correlations. The most consistent approach to
decoupling the BBGKY hierarchy was developed by N. Bogoliubov \cite{Bogoliubov
(1946)}. He suggested the principle of weakening of initial correlations which
implies that at a large enough time, $t-t_{0}\gg t_{cor}$ ($t_{cor}$ is the
correlation time caused by inter-particle interaction), \textbf{all} initial
correlations damp and, as a consequence of that, the time-dependence of
multi-particle distribution functions is completely defined by the
time-dependence of a one-particle distribution function (Bogoliubov's ansatz).
The latter assumes an existence of the time interval
\begin{equation}
t_{cor}\ll t-t_{0}\ll t_{rel}\label{0}%
\end{equation}
($t_{rel}$ is the relaxation time for a one-particle distribution function)
and leads to an approximate (valid only on the pointed above large timescale)
conversion of the inhomogeneous (including two-particle correlations) equation
for a one-particle distribution function of the BBGKY hierarchy into a
homogeneous one. By employing this approach, Bogoliubov was successful in
deriving the kinetic (Markovian) equation for a one-particle distribution
function, particularly the Boltzmann equation, which describes the
time-evolution with the characteristic time $t_{rel}$ but is unable to account
for the initial stage of the evolution, $t_{0}\leqslant t\leqslant t_{cor}$.

The Bogoliubov principle of weakening of initial correlations is introduced as
a following boundary (initial) condition to the BBGKY chain (e.g. for the
classical physics case)
\begin{align}
\lim\limits_{t-t_{0}\rightarrow\infty}U(t,t_{0})\left[  F_{N}(x_{1}%
,\ldots,x_{N},t_{0})-f_{r}(x_{1},\ldots,x_{N},t_{0})\right]   &
=0,\nonumber\\
f_{r}(x_{1},\ldots,x_{N},t)  &  =%
{\displaystyle\prod\limits_{i=1}^{N}}
F_{1}(x_{i},t), \label{0a}%
\end{align}
where $F_{N}(x_{1},\ldots,x_{N},t_{0})$ is an $N$-particle distribution
function at $t=t_{0}$, $x_{i}=(\mathbf{r}_{i},\mathbf{p}_{i})$ is a coordinate
of the $i$-particle in a phase space and $U(t,t_{0})$ is the operator defining
the evolution of all coordinates $x_{i}$ (back) in time (see (\ref{6})). The
first condition (\ref{0}) implies that at $t-t_{0}\gg t_{cor}$ all
correlations between the particles vanish and the second inequality in
(\ref{0}) $t-t_{0}\ll t_{rel}$ allows to neglect (in the first approximation
in $n$) the time-retardation and substitute in (\ref{0a}) $U(t,t_{0}%
)f_{r}(t_{0})$ with $U(t,t_{0})f_{r}(t)$. Thus, the decoupling of the BBGKY
chain is a nonlinear procedure (\ref{0a}) which results in the closed
nonlinear (like Boltzmann's) equation for a one-particle distribution function
$F_{1}(x_{i},t)$ and introduces an irreversibility into the
Liouville-von-Neumann equation.

It should be noted, however, that actually only the small-scale initial
correlations with $t_{cor}\ll t_{rel}$ may be disregarded on the kinetic
timescale. Trying to get the collision integral in the kinetic equation with
account for collisions of three and more particles (the terms of the second
and higher order in $n$) one should deal with the divergencies caused,
particularly, by the large-scale initial correlations with $t_{cor}\gtrsim
t_{rel}$ (see, e.g.\cite{Dorfman and Cohen (1967)} and \cite{Ferziger and
Kaper (1972)}). Thus, the correct expansion of the collision integral in
powers of $n$ and accounting for kinetic fluctuations are impossible if one
does not account for initial correlations with $t_{cor}\gtrsim t_{rel}$ and
$l_{cor}\gtrsim l$ ($l_{cor}=\overline{v}t_{cor}$, $l=\overline{v}t_{rel}$,
$\overline{v}$ is the characteristic (mean) particle's velocity and $l$ is the
particle's mean-free path) \cite{Klimontovich (1982)}.

In another approach, leading to the so called generalized master equation
(GME) \cite{Nakajima (1958), Zwanzig (1960)} and \cite{Prigogine (1962)}, an
$N$-particle distribution function (or statistical operator) is divided by
means of time-independent projection operators $P$ and $Q$ into the relevant
and irrelevant parts%
\begin{align}
F_{N}(x_{1},\ldots,x_{N},t)  &  =f_{r}(x_{1},\ldots,x_{N},t)+f_{i}%
(x_{1},\ldots,x_{N},t),\nonumber\\
f_{r}(x_{1},\ldots,x_{N},t)  &  =PF_{N}(x_{1},\ldots,x_{N},t),f_{i}%
(x_{1},\ldots,x_{N},t)=QF_{N}(x_{1},\ldots,x_{N},t),\nonumber\\
P+Q  &  =1. \label{0b}%
\end{align}
Note, that the relevant and irrelevant parts depend on coordinates of all $N$
particles in contrast to the reduced distribution functions (like $F_{1}%
(x_{i})$). The relevant part, which is mainly of interest, is, as a rule, a
vacuum (slowly changing) part of a distribution function (statistical
operator), i.e. the part with no correlations (like $f_{r}(x_{1},\ldots
,x_{N},t)$ in (\ref{0a})). Applying the projectors $P$ and $Q$ to the
Liouville-von-Neumann equation (see (\ref{1})), one arrives at the GME, which
is the exact inhomogeneous non-Markovian equation for the relevant part of a
distribution function with a source (irrelevant part) containing all
many-particle correlations at the initial moment $t_{0}$. It should be
underlined, that the time-independent (linear) projection operators $P$ and
$Q$, used in this approach, commute with the time-derivative operator in the
Liouville-von-Neumann equation for the distribution function (statistical
operator) of the whole system. Therefore, the procedure (\ref{0b}) is a linear
one and transforms the linear Liouville-von-Neumann equation into the linear
GME. To exclude initial many-particle correlations (a source), Bogoliubov's
principle of weakening of initial correlations or simply the factorizing
initial conditions (random phase approximation (RPA) which is incorrect in
principle \cite{van Kampen (2004)}) are commonly employed, and it results in
the approximate closed linear homogeneous GME for the relevant part of a
distribution function (statistical operator). The latter is then used for
obtaining a linear evolution equation for an interesting reduced distribution
function. \ 

The abovesaid means that the proper treating of initial correlations is a key
to obtaining the closed kinetic equations. To include initial correlations
into consideration, the method (based on the conventional projection operator
technique) turning the conventional linear inhomogeneous generalized master
equations (GMEs) into the homogeneous form has been proposed recently
\cite{Los (2001), Los (2005)}. By introducing an additional identity for the
irrelevant initial condition term, it has become possible to express this term
through the relevant part of a distribution function. Thus, the irrelevant
initial condition (inhomogeneous) terms in time-convolution (non-Markovian)
GME (TC-GME) \cite{Nakajima (1958), Zwanzig (1960), Prigogine (1962)} and
time-convolutionless GME (TCL-GME) \cite{Shibata et al (1977), Shibata and
Arimitsu (1980)}, caused by the correlations at an initial (at $t=t_{0}$)
state of the whole system, have been transferred to the memory kernels of GMEs
governing the evolution of the relevant part of a distribution function. That
resulted in exact time-convolution and time-convolutionless homogeneous
generalized master equation (TC-HGME and TCL-HGME) for the relevant part of a
distribution function (statistical operator) of a many-particle system which
account for dynamics of initial correlations through the modified memory
kernels. These equations, which are obtained with no loss of information,
describe the evolution (influenced by initial correlations) of the relevant
part of a distribution function at all timescales including the initial one,
$t-t_{0}\lesssim t_{cor}$, when the transition from the reversible evolution
to the irreversible kinetic behavior (caused by the stochastic instability of
the system's dynamics) is expected to occur.

The obtained linear time-convolution homogeneous generalized master equation
(TC-HGME) has been tested for the space-homogeneous dilute gas of classical
\cite{Los (2001)} and quantum \cite{Los (2005)} particles. However, in order
to get from the actually linear TC-HGME the desired nonlinear equations
(particularly the Boltzmann equation) describing the evolution of classical or
quantum gas of particles, the additional approximation disregarding the change
in time of a one-particle distribution function (statistical operator) during
collisions should be performed. This is because the time-independent
projection operator (for, e.g. the classical particles) \cite{Los (2001)}%
\begin{equation}
P=[%
{\displaystyle\prod\limits_{i=2}^{N}}
F_{1}(x_{i},t_{0})]\frac{1}{V^{N-1}}\int dx_{2}\ldots\int dx_{N} \label{0c}%
\end{equation}
($V$ is the volume of the system) leads to the relevant part of an
$N$-particle distribution function%
\begin{equation}
f_{r}(x_{1},\ldots,x_{N},t)=[%
{\displaystyle\prod\limits_{i=2}^{N}}
F_{1}(x_{i},t_{0})]F_{1}(x_{1},t), \label{0d}%
\end{equation}
which is linear with respect to the time-dependent one-particle distribution
function of interest $F_{1}(x_{1},t)$ ($F_{N}$ satisfies the normalization condition

$\frac{1}{V^{N}}\int dx_{1}\ldots\int dx_{N}F_{N}(x_{1},\ldots,x_{N},t)=1$).
Then, in order to get from obtained linear equation for $F_{1}(x_{1},t)$, the
nonlinear one, $F_{1}(x_{i},t_{0})$ may be approximated by $F_{1}(x_{i},t)$
\textbf{if }$t-t_{0}\ll t_{rel}$ (see (\ref{0})). Thus, in this approach we
are not restricted by the first inequality (\ref{0}), $t_{cor}\ll t-t_{0}$,
but still need its second part. It is difficult to imagine how to get a
nonlinear symmetric in time relevant distribution function $f_{r}(x_{1}%
,\ldots,x_{N},t)$ (\ref{0a}) by applying the time-independent projection
operator $P$ to the $N$-particle distribution function.

In this paper we will obtain a new class of evolution equations for the
relevant part of a distribution function (statistical operator) suitable for
deriving both the nonlinear and linear evolution equations for the reduced
distribution functions (statistical operators) of interest and which are valid
at all timescales. It will be exact \textbf{nonlinear }inhomogeneous and
homogeneous generalized master equations deduced from the
Liouville-von-Neumann equation with the help of (generally) nonlinear
time-dependent operator $P(t)$ converting the many-particle distribution
function (statistical operator) into the appropriate relevant form. The
time-dependent operator $P(t)$ is not generally a projection operator. For the
case of a time-independent linear projection operator $P$ the obtained
equations reduce to the linear conventional GME and obtained recently HGME. In
contrast to these linear equations, the obtained nonlinear equations are more
general and, therefore, more convenient, particularly, for studying the
space-inhomogeneous nonideal gas of particles when the space-nonlocality and
time-retardation are essential (see, e.g. \cite{Klimontovich (1982)}). In
addition, the initial correlations may be exactly included into consideration
by converting inhomogeneous nonlinear GME into the homogeneous form.

At first (Sec. 2), we will derive the exact inhomogeneous nonlinear
time-convolution GME for the relevant part of a distribution function
(statistical operator) containing an irrelevant part (comprising initial
correlations) which poses a problem do deal with as in the case of
conventional linear GME. Actually, this equation provides an alternative to
the BBGKY chain. Like linear inhomogeneous GMEs, the obtained nonlinear
equation will be transformed in Sec. 4 into the homogeneous nonlinear
time-convolution GME by the method suggested earlier in \cite{Los (2001)}. The
derived exact homogeneous nonlinear equation for the relevant part of a
distribution function (statistical operator) accounts for initial correlations
by means of the modified memory kernel of inhomogeneous nonlinear equation. It
is worth noting that there is no way yet to exactly account for initial
correlations dealing with the BBGKY chain.

The obtained nonlinear GMEs will be applied to the space inhomogeneous
nonideal dilute gas of classical particles by choosing the appropriate
nonlinear operator $P(t)$. In the linear approximation in the small density
parameter (only two-particle correlations (collisions) are taken into
consideration), the inhomogeneous nonlinear GME results in the inhomogeneous
nonlinear equation for a one-particle distribution function containing the
self-consistent Vlasov term, the collision integral accounting for all space-
and time-nonlocalities and an irrelevant part (a two-particle correlation
function at the initial moment of time $t_{0}$). Accepting RPA for $t=t_{0}%
$(or Bogoliubov's principle of weakening of initial correlations for
$t_{0}\rightarrow-\infty$) and disregarding the terms accounting for the
space- and time-change of a distribution function in the collision integral on
the microscopic scales $l_{cor}$ and $t_{cor}$(which are of the small density
parameter order), we obtain the Boltzmann equation (Sec. 3).

Then, in order to include initial correlations into consideration, we apply
the derived homogeneous nonlinear GME to the same system (Sec. 5). In the
linear approximation in the density parameter we obtain the homogenous
nonlinear equation for a one-particle distribution function valid at all
timescales (no restrictions like (\ref{0}) are needed) and accounting for
influence of the dynamics of initial correlations at all stages of the
evolution process. In this linear (in $n$) approximation, the evolution of
collisions and initial correlations is governed by the exact two-particle
propagator, and in this sense the obtained equation is closed. If all initial
correlations vanish at the kinetic timescale (and all nonlocalities of the
collision integral contributing to the thermodynamic functions of the nonideal
gas are omitted), this equation switches to the nonlinear Boltzmann equation
with the self-consistent nonlinear Vlasov term. The results are summarized in
Sec. 6.

\section{Inhomogeneous nonlinear generalized master equation}

As it was already mentioned, the aim of this paper is to generalize the method
suggested in \cite{Los (2001)} and \cite{Los (2005)} for deriving the
homogeneous generalized master equations (HGMEs) retaining initial
correlations to the case when the (projection) operator selecting the relevant
part of a distribution function (statistical operator) depends on time. It
will allow for obtaining more general equations (inhomogeneous and homogeneous
nonlinear generalized master equations) which, particularly, are more flexible
in selecting the suitable for the problem under consideration relevant
distribution function (statistical operator) describing the evolution of a
(sub)system of interest. We will also generalize our approach to the case of
the time-dependent external forces.

We start with the Liouville-von-Neumann equation describing the evolution of a
distribution function (statistical operator) $F(t)$ of a total system under
consideration%
\begin{equation}
\frac{\partial}{\partial t}F(t)=L(t)F(t), \label{1}%
\end{equation}
where, $L(t)$ is the Liouville (evolution) operator acting on $F(t)$ and
depending, generally, on time. For example, in the case, when $F(t)$ is
related to the system of $N$ classical particles and depends on their
coordinates and momenta $x_{i}=(\mathbf{r}_{i},\mathbf{p}_{i})$ ($i=1,\ldots
,N$),
\begin{equation}
L(t)F(t)=\left\{  H(t),F(t)\right\}  _{P}=\sum_{i=1}^{N}\left[  \frac{\partial
H(t)}{\partial\mathbf{r}_{i}}\frac{\partial F(t)}{\partial\mathbf{p}_{i}%
}-\frac{\partial H(t)}{\partial\mathbf{p}_{i}}\frac{\partial F(t)}%
{\partial\mathbf{r}_{i}}\right]  , \label{2}%
\end{equation}
where $H(t)$ is the corresponding time-dependent Hamilton function of all
$x_{i}$. For the quantum physics case, $L(t)$ is the superoperator acting on a
statistical operator as%
\begin{equation}
L(t)F(t)=\frac{1}{i\hbar}\left[  H(t),F(t)\right]  , \label{3}%
\end{equation}
where $[,]$ is a commutator and $H(t)$ is the energy operator (Hamiltonian).
The distribution function (or statistical operator) is subjected to the
normalization condition
\begin{equation}
\int dx_{1}\ldots\int dx_{N}F(x_{1},\ldots x_{N},t)=1 \label{4}%
\end{equation}
or%
\begin{equation}
TrF(t)=1. \label{5}%
\end{equation}

The formal solution to equation (\ref{1}) may be written as%
\begin{equation}
F(t)=U(t,t_{0})F(t_{0}),U(t,t_{0})=T\exp\left[
{\textstyle\int\limits_{t_{0}}^{t}}
dsL(s)\right]  , \label{6}%
\end{equation}
where $T$ denotes the chronological time-ordering operator which orders the
product of time-dependent operators such that their time-arguments increase
from right to left and $F(t_{0})$ is the value of a distribution function
(statistical operator) at some initial time $t_{0}$.

Let us introduce the generally time-dependent operator $P(t)$ converting the
distribution function $F(t)$ into a relevant distribution function $f_{r}(t)$,
which, as a rule, is a vacuum (with no correlations) many-particle
distribution function, i.e. a product of the reduced distribution functions
which are necessary and sufficient for describing the evolution of the
measurable (statistical expectation) values.

However, the problem is to obtain the closed evolution equation for such
reduced distribution functions in view of the existence of correlations
between the elements of the considered system including initial correlations
at the initial time $t_{0}$.

Thus, we define the relevant distribution function (statistical operator) as%
\begin{equation}
f_{r}(t)=P(t)F(t)=P(t)U(t,t_{0})F(t_{0}). \label{7}%
\end{equation}
Note, that generally the operator $P(t)$ is not a projection operator.
Moreover, the action of \ $P(t)$ on $F(t)$ is not a linear operation anymore
(in contrast to applying the conventional time-independent projection operator
\cite{Zwanzig (1960)}) because $P(t)$ may depend on the time-dependent
distribution function of interest (see below). Therefore, by applying $P(t)$
to the linear Liouville equation (\ref{1}), one will generally obtain a
nonlinear equation. Because $P(t^{^{\prime}})F(t)\neq f_{r}(t)$ (at
$t^{^{\prime}}\neq t$), operator $P(t)$ does not commute with the derivative
$\frac{\partial}{\partial t}$ (which is the case for the linear operators).

Thus, using the Liouville-von-Neumann equation (\ref{1}), the equation of
motion for $f_{r}(t)$ (\ref{7}) may be written as%
\begin{equation}
\frac{\partial f_{r}(t)}{\partial t}=\left[  \frac{\partial P(t)}{\partial
t}+P(t)L(t)\right]  \left[  f_{r}(t)+f_{i}(t)\right]  . \label{8}%
\end{equation}
Here, $f_{i}(t)=\left[  F(t)-f_{r}(t)\right]  $, $\frac{\partial
P(t)}{\partial t}$ denotes the operator obtained by taking the derivative of
the operator $P(t)$, i.e., we assume that the operator $\frac{\partial
P(t)}{\partial t}$ exists. To make sense of $\frac{\partial P(t)}{\partial t}%
$, one may define $P(t)$ as
\begin{equation}
P(t)=C(t)D, \label{8a}%
\end{equation}
where $C(t)$ is a well defined (operator) function of time and $D$ is a
time-independent (super)operator acting on $F(t)$ and integrating off all
unnecessary variables in the spirit of the reduced description method (more
details on the reasonable specification of $P(t)$ will be given below).

As it follows from (\ref{8}), the equation for the irrelevant part of a
distribution function $f_{i}(t)=F(t)-f_{r}(t)$ may be obtained by using
equations (\ref{1}) and (\ref{8}) and splitting $F(t)$ into the relevant and
irrelevant parts. Thus,%

\begin{equation}
\frac{\partial f_{i}(t)}{\partial t}=\left[  Q(t)L(t)-\frac{\partial
P(t)}{\partial t}\right]  \left[  f_{i}(t)+f_{r}(t)\right]  , \label{9}%
\end{equation}
where $Q(t)=1-P(t)$ and, therefore, $f_{i}(t)=Q(t)F(t)$.

The formal solution to equation (\ref{9}) is%
\begin{equation}
f_{i}(t)=%
{\textstyle\int\limits_{t_{0}}^{t}}
dt^{^{\prime}}S(t,t^{^{\prime}})\left[  Q(t^{^{\prime}})L(t^{^{\prime}}%
)-\frac{\partial P(t^{^{\prime}})}{\partial t^{^{\prime}}}\right]
f_{r}(t^{^{\prime}})+S(t,t_{0})f_{i}(t_{0}), \label{10}%
\end{equation}
where
\begin{equation}
S(t,t_{0})=T\exp\left\{
{\textstyle\int\limits_{t_{0}}^{t}}
ds\left[  Q(s)L(s)-\frac{\partial P(s)}{\partial s}\right]  \right\}  .
\label{11}%
\end{equation}
The latter operator may be presented as a following series%
\begin{align}
S(t,t_{0})  &  =1+%
{\textstyle\int\limits_{t_{0}}^{t}}
dt_{1}\left[  Q(t_{1})L(t_{1})-\frac{\partial P(t_{1})}{\partial t_{1}}\right]
\nonumber\\
&  +%
{\textstyle\int\limits_{t_{0}}^{t}}
dt_{1}%
{\textstyle\int\limits_{t_{1}}^{t}}
dt_{2}\left[  Q(t_{2})L(t_{2})-\frac{\partial P(t_{2})}{\partial t_{2}}\right]
\nonumber\\
&  \times\left[  Q(t_{1})L(t_{1})-\frac{\partial P(t_{1})}{\partial t_{1}%
}\right]  +\ldots\label{11a}%
\end{align}
Note, that if the operators $P$ and $L$ do not depend on time, then
$S(t,t_{0})=\exp\left[  QL(t-t_{0})\right]  $.

Inserting (\ref{10}) into (\ref{8}), we get%
\begin{align}
\frac{\partial f_{r}(t)}{\partial t}  &  =\left[  P(t)L(t)+\frac{\partial
P(t)}{\partial t}\right]  \{f_{r}(t)\nonumber\\
&  +%
{\textstyle\int\limits_{t_{0}}^{t}}
dt^{^{\prime}}S(t,t^{^{\prime}})\left[  Q(t^{^{\prime}})L(t^{^{\prime}}%
)-\frac{\partial P(t^{^{\prime}})}{\partial t^{^{\prime}}}\right]
f_{r}(t^{^{\prime}})\nonumber\\
&  +S(t,t_{0})f_{i}(t_{0})\}. \label{12}%
\end{align}

Equation (\ref{12}) is the generalization of the Nakajima-Zwanzig
time-convolution generalized master equation (TC-GME) (see \cite{Nakajima
(1958), Zwanzig (1960), Prigogine (1962)}) for the relevant part of a
distribution function (statistical operator) to the case of time-dependent
operators $P$ and $Q$. As it has been already noted, $P(t)$ and $Q(t)$ are not
generally the projection operators in the usual sense and in our derivation of
(\ref{12}) we have never used such a property of the projection operators as
$P^{2}=P$, $Q^{2}=Q$ (for $P(t)P(t^{^{\prime}})$ and $Q(t)Q(t^{^{\prime}})$
such a property holds only for $t=t^{^{\prime}}$, as it will be seen below).
The obtained equation, like conventional TC-GME, is an exact inhomogeneous
integra-differential equation for the relevant part of a distribution function
(statistical operator) containing the irrelevant part of a distribution
function (a source) at initial moment of time $f_{i}(t_{0})=F(t_{0}%
)-f_{r}(t_{0})$. But, equation (\ref{12}) is generally the nonlinear equation
for a reduced distribution function of interest (e.g. for a one-particle
distribution function considered in the next section) and, therefore, is also
convenient for studying the evolution of many-particle systems, which are
described by the nonlinear Boltzmann-type equations (in the kinetic regime)
and where the nonlinearity is caused by the collisions between particles.

For time-independent $P$ and $Q$ equation (\ref{12}) reduces to the
conventional linear TC-GME. This type of equations is more suitable for
studying, e.g. the evolution of a subsystem interacting with a large system in
the thermal equilibrium state (a thermostat).

To simplify equation (\ref{12}) and make sense of the derivative
$\frac{\partial P(t)}{\partial t}$, let us consider the operator $P(t)$ of the
form (\ref{8a}), $P(t)=C(t)D$. This representation of $P(t)$ is suitable for
studying many cases of interest (see the next section). Also, splitting a
distribution function (statistical operator) into the relevant $P(t)F(t)$ and
irrelevant $Q(t)F(t)$ parts makes sense if the operator $P(t)$ satisfies the
following relation \ \ \ \ \
\begin{equation}
P(t)P(t)=P(t) \label{12a}%
\end{equation}
for any $t$. Given the form of $P(t)$, $P(t)=C(t)D$, equation (\ref{12a})
entails
\begin{equation}
DC(t)=1 \label{12b}%
\end{equation}
for any time moment $t$. Equation (\ref{12b}) generally represents the
normalization condition for the distribution function(s) making up $C(t)$ (see
(\ref{4}), (\ref{5}) and (\ref{17}) below) and implies that $C(t)$ depends on
the variables which are removed from $F(t)$ by operator $D$ ($f_{r}%
(t)=P(t)F(t)$ depends on the complete set of the variables of a distribution
function $F(t)$ as indicated by (\ref{0b})). From the property (\ref{12b}) we
have
\begin{align}
P(t)P(t^{^{\prime}})  &  =P(t),Q(t)Q(t^{^{\prime}})=Q(t^{^{\prime}%
}),P(t)Q(t^{^{\prime}})=0,\nonumber\\
Q(t^{^{\prime}})P(t)  &  =P(t)-P(t^{^{\prime}}). \label{12c}%
\end{align}
The condition (\ref{12b}) also leads to the following relations for the
time-derivative of $P(t)$
\begin{equation}
D\frac{\partial P(t)}{\partial t}=D\frac{\partial C(t)}{\partial t}D=0,
\label{12d}%
\end{equation}
where we assumed that the operators $D$ and $\partial/\partial t$ commute ($D$
does not depend on time).

Let us also introduce the reduced distribution function (statistical operator)
$f_{red}(t)$%
\begin{equation}
f_{red}(t)=DF(t)=Df_{r}(t), \label{12e}%
\end{equation}
which is actually needed for calculation of the expectation values of interest
in the non-equilibrium state. Also, due to (\ref{12b}),%
\begin{equation}
D\frac{\partial f_{r}(t)}{\partial t}=\frac{\partial f_{red}(t)}{\partial
t},Df_{i}(t)=DQ(t)F(t)=0. \label{12f}%
\end{equation}

Thus, by applying the operator $D$ to equation (\ref{12}) from the left, we
get%
\begin{align}
\frac{\partial f_{red}(t)}{\partial t}  &  =DL(t)\{f_{r}(t)+%
{\textstyle\int\limits_{t_{0}}^{t}}
dt^{^{\prime}}S(t,t^{^{\prime}})\left[  Q(t^{^{\prime}})L(t^{^{\prime}}%
)-\frac{\partial P(t^{^{\prime}})}{\partial t^{^{\prime}}}\right]
f_{r}(t^{^{\prime}})\nonumber\\
&  +S(t,t_{0})f_{i}(t_{0})\}, \label{12g}%
\end{align}
where (\ref{12b}), (\ref{12d}) and (\ref{12f}) have been used.

Without loss of generality the Hamiltonian of a system may be split into two
terms as $H(t)=H^{0}(t)+H^{^{\prime}}(t)$ ($L(t)=L^{0}(t)+L^{^{\prime}}(t)$,
correspondingly), where $H^{0}$($L^{0}$) is related to the energy of the
noninteracting particles (or subsystems) and $H^{^{\prime}}$($L^{^{\prime}} $)
describes the interaction between particles (subsystems). Then, except
satisfying the condition (\ref{12b}), the operator $D$, selecting the reduced
distribution function, should commute with $L^{0}$in the following manner%
\begin{equation}
DL^{0}(t)=L_{red}^{0}(t)D, \label{12h}%
\end{equation}
where $L_{red}^{0}(t)$ is the reduced Liouvillian $L^{0}$ depending only on
the variables of the reduced distribution function (statistical operator)
$f_{red}(t)$, and (\ref{12h}) means that both sides of this relation act on
the arbitrary function (operator) defined on the phase (Hilbert) space. Taking
into account that $C(t^{^{\prime}})L_{red}^{0}(t)=L_{red}^{0}(t)C(t^{^{\prime
}})$ (because $C(t^{^{\prime}})$ depends on the set of variables that are
removed from $F(t)$ by $D$ and $L_{red}^{0}(t)$ does not depend on them),
equation (\ref{12h}) may be written in the more general form%
\begin{equation}
P(t^{^{\prime}})L^{0}(t)=L_{red}^{0}(t)P(t^{^{\prime}}). \label{12h'}%
\end{equation}
The commutation relation (\ref{12h}) follows from a self-consistency reason,
because if $L^{^{\prime}}(t)=0$, the evolution of a "free" (noninteracting
with the rest of a system) subsystem should be described in equation
(\ref{12g}) only by $L_{red}^{0}(t)$ ($DL^{0}(t)f_{r}(t)=L_{red}^{0}%
(t)f_{red}(t)$) (see also \cite{Bogoliubov (1946)}). That this is the case may
be seen from the relations%
\begin{align}
DL^{0}(t)S(t,t_{1})Q(t_{2})  &  =0,\nonumber\\
DL^{0}S(t,t_{1})\frac{\partial P(t_{2})}{\partial t_{2}}  &  =0, \label{12i}%
\end{align}
which follow from (\ref{12h}), (\ref{11}), (\ref{12d}) and (\ref{12f}). Thus,
all terms in (\ref{12g}) are proportional to $L^{^{\prime}}(t)$ but the first
one ($L_{red}^{0}(t)f_{red}(t)$).

In the same way (using (\ref{12h}), (\ref{11}) and relations that follow from
(\ref{12b})) the following equations may be proved%
\begin{align}
S(t,t_{1})Q(t_{2})  &  =\overline{U}(t,t_{1})Q(t_{2}),\nonumber\\
S(t,t_{1})\frac{\partial P(t_{2})}{\partial t_{2}}  &  =\overline{U}%
(t,t_{1})\frac{\partial P(t_{2})}{\partial t_{2}}, \label{12j}%
\end{align}
where
\begin{equation}
\overline{U}(t,t_{1})=T\exp\left\{
{\textstyle\int\limits_{t_{1}}^{t}}
ds\left[  L^{0}(s)+Q(s)L^{^{\prime}}\right]  \right\}  . \label{12k}%
\end{equation}
Thus, making use of (\ref{12i}) and (\ref{12j}) in (\ref{12g}), equation
(\ref{12}) may be finally rewritten as the following equation for the relevant
part of a distribution function (statistical operator)%
\begin{align}
\frac{\partial Df_{r}(t)}{\partial t}  &  =DL(t)f_{r}(t)\nonumber\\
&  +DL^{^{\prime}}(t)%
{\textstyle\int\limits_{t_{0}}^{t}}
dt^{^{\prime}}\overline{U}(t,t^{^{\prime}})\left[  Q(t^{^{\prime}%
})L(t^{^{\prime}})-\frac{\partial P(t^{^{\prime}})}{\partial t^{^{\prime}}%
}\right]  f_{r}(t^{^{\prime}})\nonumber\\
&  +DL^{^{\prime}}(t)\overline{U}(t,t_{0})f_{i}(t_{0}), \label{12l}%
\end{align}
where $Df_{r}(t)=f_{red}(t)$ and (\ref{12h}) holds.

The fulfillment of the basic conditions (\ref{12b}) and (\ref{12h}) will be
demonstrated in the next section for a gas of interacting classical particles.

\section{Dilute inhomogeneous gas of classical particles}

Let us apply equation (\ref{12l}) to the case of a gas of $N$ ($N\gg1$)
identical classical particles. For such a system the Liouville operator $L$
may be presented as%
\begin{align}
L  &  =L^{0}+L^{^{\prime}},\nonumber\\
L^{0}  &  =%
{\textstyle\sum\limits_{i=1}^{N}}
L_{i}^{0},L_{i}^{0}=-\mathbf{v}_{i}\cdot\mathbf{\nabla}_{i},\mathbf{v}%
_{i}=\frac{\mathbf{p}_{i}}{m}\text{ },\mathbf{\nabla}_{i}=\frac{\partial
}{\partial\mathbf{x}_{i}}\text{ },\nonumber\\
L^{^{\prime}}  &  =\sum\limits_{i<j=1}^{N}L_{ij}^{^{\prime}},L_{ij}^{^{\prime
}}=(\mathbf{\nabla}_{i}V_{ij})\cdot(\frac{\partial}{\partial\mathbf{p}_{i}%
}-\frac{\partial}{\partial\mathbf{p}_{j}}), \label{13}%
\end{align}
where $L^{0}$ corresponds to the kinetic energy $H^{0}=%
{\textstyle\sum\limits_{i=1}^{N}}
\mathbf{p}_{i}^{2}/2m$ of particles with the momenta $\mathbf{p}_{i}$ and mass
$m$, while $L^{^{\prime}}$corresponds to interaction between particles
$H^{^{\prime}}=%
{\textstyle\sum\limits_{i<j=1}^{N}}
V_{ij}$ with the pair potential $V_{ij}=V(|\mathbf{x}_{i}-\mathbf{x}_{j}|)$.
Thus, in the considered case $L$ does not depend on time. We do not suppose
here that the interaction is weak and assume that all necessary requirements
to the properties of forces, by which the particles interact with each other,
are met (particularly, the bound states are not formed).

We will also assume, as usual, that all functions $\Phi(x_{1},\ldots,x_{N}) $,
defined on the phase space, and their derivatives vanish at the boundaries of
the configurational space and at $\mathbf{p}_{i}=\pm\infty$. These boundary
conditions and the explicit form of the Liouville operators (\ref{13}) lead to
the following relations
\begin{align}
\int dx_{i}L_{i}^{0}\Phi(x_{1},\ldots,x_{N},t)  &  =0,\nonumber\\
\int dx_{i}\int dx_{j}L_{ij}^{^{\prime}}\Phi(x_{1},\ldots,x_{N},t)  &  =0.
\label{13a}%
\end{align}

We are looking for an evolution equation for a one-particle distribution
function
\begin{equation}
F_{1}(x_{i},t)=V\int dx_{1}\cdots\int dx_{i-1}\int dx_{i+1}\cdots\int
dx_{N}F(x_{1},\ldots,x_{N},t). \label{14}%
\end{equation}
To this end, it is convenient to define the operator $P(t)$ as%
\begin{equation}
P(t)=C(t)D,C(t)=\left[  \prod\limits_{i=2}^{N}F_{1}(x_{i},t)\right]
,D=\frac{1}{V^{N-1}}\int dx_{2}\cdots\int dx_{N}, \label{15}%
\end{equation}
where $V$ is the volume of the system.

Generally, the $s$-particle ($s\leqslant N$) distribution function is defined
as%
\begin{equation}
F_{s}(x_{1},\ldots,x_{s},t)=V^{s}\int dx_{s+1}\ldots\int dx_{N}F(x_{1}%
,\ldots,x_{N},t) \label{16}%
\end{equation}
and satisfies the normalization condition (see (\ref{4}))%
\begin{equation}
\int dx_{1}\ldots\int dx_{s}F_{s}(x_{1},\ldots,x_{s},t)=V^{s}. \label{17}%
\end{equation}
Because the distribution function (\ref{16}) for $s=N$, $F_{N}(t)=V^{N}F(t)$
(hereinafter the dependence on $x_{i}$ is often suppressed for brevity),
satisfies the same Liouville equation (\ref{1}), we may apply the projection
operator (\ref{15}) to $F_{N}(t)$ (instead of $F(t)$) and obtain the following
relevant part of the distribution function%
\begin{equation}
f_{r}(t)=P(t)F_{N}(t)=%
{\textstyle\prod\limits_{i=1}^{N}}
F_{1}(x_{i},t). \label{18}%
\end{equation}
From the definition (\ref{18}) one can see an advantage of the time-dependent
operator $P(t)$: by applying it to an $N$-particle distribution function
$F_{N}(t)$, it is possible to define the relevant part (\ref{18}) of a
distribution function, which looks more natural and symmetric than (\ref{0d})
used in \cite{Los (2001)} and obtained with the help of the time-independent
projection operator (\ref{0c}) (we had to use the time-independent projection
operator because the TC-GME considered in \cite{Los (2001)} is valid only for
such operators). Moreover, the relevant part (\ref{18}) has a clear physical
meaning because it represents a slow changing vacuum (with no correlations)
part of an $N$-particle distribution function which is mostly of interest.

Thus, the irrelevant part%
\begin{equation}
f_{i}(t)=Q(t)F_{N}(t)=F_{N}(t)-%
{\textstyle\prod\limits_{i=1}^{N}}
F_{1}(x_{i},t) \label{18a}%
\end{equation}
describes all inter-particle correlations, particularly the initial ones
(existing at initial moment of time $t_{0}$), and makes up the inhomogeneous
term of (\ref{12l}). The irrelevant part of the $N$-particle distribution
function at $t_{0}$, $f_{i}(t_{0})=F_{N}(t_{0})-f_{r}(t_{0})$, may be always
represented (as well as at any other time $t$) by the following cluster
expansion%
\begin{equation}
f_{i}(t_{0})=\sum\limits_{i<j=1}^{N}g_{2}(x_{i},x_{j})\prod\limits_{k\neq
i,j}^{N-2}F_{1}(x_{k})+\sum\limits_{i<j<k=1}^{N}g_{3}(x_{i},x_{j},x_{k}%
)\prod\limits_{l\neq i,j,k}^{N-3}F_{1}(x_{l})+\ldots, \label{18b}%
\end{equation}
where $\prod\limits_{k\neq i,j}^{N-2}F_{1}(x_{k})$ and $\prod\limits_{l\neq
i,j,k}^{N-3}F_{1}(x_{l})$ stand for the products of $N-2$ and $N-3$
one-particle distribution functions with $k\neq i,j$ and $l\neq i,j,k$,
respectively, whereas $g_{2}(x_{i},x_{j})$ and $g_{3}(x_{i},x_{j},x_{k})$ are
the irreducible two-particle and three-particle correlation functions (further
terms in (\ref{18b}) are defined in the same way). Here and further, the
one-particle distribution and correlation functions with the suppressed time
argument denote the functions taken at $t=t_{0}$ (e.g., $F_{1}(x_{k}%
)=F_{1}(x_{k},t_{0})$).

Using the normalization condition (\ref{17}) for a one-particle distribution
function, $\int F_{1}(x_{i},t)dx_{i}=V$, it is easy to see, that by applying
the integral operator $D$ to $C(t)$ (both defined by (\ref{15})) we obtain
$DC(t)=1$, i.e. the condition (\ref{12b}) and the consequences of this
condition given by (\ref{12c}) and (\ref{12d}) hold. Also, in this case
\begin{align}
f_{red}(t)  &  =Df_{r}(t)=F_{1}(x_{1},t),\nonumber\\
DQ(t)  &  =0,Df_{i}(t)=0 \label{19}%
\end{align}
\bigskip in accordance with (\ref{12e}) and (\ref{12f}).

The commutation rules (\ref{12h}) and (\ref{12h'}) also follow from
(\ref{13}), (\ref{13a}) and (\ref{15})%
\begin{align}
DL^{0}  &  =L_{1}^{0}D\nonumber\\
P(t)L^{0}  &  =L_{1}^{0}P(t), \label{20}%
\end{align}
i.e. in the case under consideration $L_{red}^{0}=L_{1}^{0}$. Therefore, the
identities (\ref{12i}), (\ref{12j}) and (\ref{12k}) also hold, where $L^{0}$
and $L^{^{\prime}}$ do not depend on time in the considered case.

Making use of (\ref{13a}), (\ref{19}) and (\ref{20}), equation (\ref{12l}) may
be presented in the case under consideration as%

\begin{align}
\frac{\partial F_{1}(x_{1},t)}{\partial t}  &  =L_{1}^{0}F_{1}(x_{1},t)+n\int
dx_{2}L_{12}^{^{\prime}}F_{1}(x_{2},t)F_{1}(x_{1},t)\nonumber\\
&  +DL^{^{\prime}}%
{\textstyle\int\limits_{t_{0}}^{t}}
dt^{^{\prime}}\overline{U}(t,t^{^{\prime}})\left[  (L^{0}-L_{1}^{0}%
)+Q(t^{^{\prime}})L^{^{\prime}}-\frac{\partial P(t^{^{\prime}})}{\partial
t^{^{\prime}}}\right]  f_{r}(t^{^{\prime}})\nonumber\\
&  +DL^{^{\prime}}\overline{U}(t,t_{0})f_{i}(t_{0}), \label{22}%
\end{align}
where $n=N/V$ is the density of particles, $D$, $f_{r}(t)$ and $f_{i}(t_{0})$
are defined by (\ref{15}), (\ref{18}) and (\ref{18b}), respectively. The
evolution operator (\ref{12k}) may be presented by the series like (\ref{11a})%
\begin{align}
\overline{U}(t,t^{^{\prime}})  &  =1+%
{\textstyle\int\limits_{t^{^{\prime}}}^{t}}
dt_{1}[L^{0}+Q(t_{1})L^{^{\prime}}]\nonumber\\
&  +%
{\textstyle\int\limits_{t^{^{\prime}}}^{t}}
dt_{1}%
{\textstyle\int\limits_{t_{1}}^{t}}
dt_{2}[L^{0}+Q(t_{2})L^{^{\prime}}][L^{0}+Q(t_{1})L^{^{\prime}}]+\ldots
\label{23}%
\end{align}

Let us consider equation (\ref{22}) in the first approximation in the
particles' density $n$. The corresponding dimensionless small parameter of a
perturbation expansion is
\begin{equation}
\gamma=r_{0}^{3}n\ll1, \label{24}%
\end{equation}
where $r_{0}$ is the effective radius of an inter-particle interaction. From
the second equation (\ref{13a}) it is easy to see, that all terms in the
expansion (\ref{23}) containing $P(t)L^{^{\prime}}$result in the expressions,
proportional at least to the first power of $n$ (like the second term in the
right-hand side of (\ref{22})). Thus, in the first approximation in $n$ all
terms of (\ref{23}) with $P(t)L^{^{\prime}}$ may be disregarded, because the
terms of equation (\ref{22}) containing $\overline{U}(t,t^{^{\prime}})$ are
already of the first order in $n$ (application of $DL^{^{\prime}}$ leads to
the expression of at least the first order in $n$). In this approximation all
terms under the integrals in (\ref{23}) do not depend on time and the operator
$\overline{U}(t,t^{^{\prime}})$ reduces to
\begin{equation}
\overline{U}(t,t^{^{\prime}})=\exp[(L^{0}+L^{^{\prime}})(t-t^{^{\prime}})].
\label{25}%
\end{equation}

Now, it is convenient to use the following expansion for the exponential
operator (\ref{25})%
\begin{equation}
e^{(L^{0}+L^{^{\prime}})(t-t^{^{\prime}})}=e^{L^{0}(t-t^{^{\prime}})}%
+\int\limits_{t^{^{\prime}}}^{t}d\theta e^{L^{0}(t-\theta)}L^{^{\prime}%
}e^{(L^{0}+L^{^{\prime}})\theta}. \label{26}%
\end{equation}
Making use of this expansion and relations (\ref{13a}), we get from (\ref{22})
the following evolution equation for a one-particle distribution function in
the linear approximation in the density parameter (\ref{24})%
\begin{align}
\frac{\partial F_{1}(x_{1},t)}{\partial t}  &  =L_{1}^{0}F_{1}(x_{1},t)+n\int
dx_{2}L_{12}^{^{\prime}}F_{1}(x_{2},t)F_{1}(x_{1},t)\nonumber\\
&  +n\int dx_{2}L_{12}^{^{\prime}}%
{\textstyle\int\limits_{0}^{t-t_{0}}}
dt_{1}e^{(L_{12}^{0}+L_{12}^{^{\prime}})t_{1}}[(L_{2}^{0}+L_{12}^{^{\prime}%
})F_{1}(x_{2},t-t_{1})F_{1}(x_{1},t-t_{1})\nonumber\\
&  +\frac{\partial F_{1}(x_{2},t-t_{1})}{\partial t_{1}}F_{1}(x_{1}%
,t-t_{1})]\nonumber\\
&  +n\int dx_{2}L_{12}^{^{\prime}}e^{(L_{12}^{0}+L_{12}^{^{\prime}})(t-t_{0}%
)}g_{2}(x_{1},x_{2}), \label{27}%
\end{align}
where $L_{12}^{0}=L_{1}^{0}+L_{2}^{0}$. Obtaining (\ref{27}), we have also
used the definition for the irrelevant part of a distribution function
(\ref{18b}) and taken into account that each additional integration over
$x_{3},\ldots$ adds an additional power of $n$ (therefore, in the linear
approximation in $n$ all formulae contain no more than one integration over
the phase space).

Equation (\ref{27}) is the main result of this section. It is necessary to
stress that equation (\ref{27}) is a \textbf{nonlinear }inhomogeneous
time-convolution (non-Markovian) master equation for a one-particle
distribution function containing initial correlations (a source). Such an
equation cannot be derived from the conventional linear time-convolution
Nakajima-Zwanzig generalized master equation (TC-GME) \cite{Nakajima (1958),
Zwanzig (1960), Prigogine (1962)}, which follows from the linear Liouville
equation with the help of the time-independent linear projection operator.
Equation (\ref{27}) is exact in the linear approximation in the small density
parameter (\ref{24}) and is valid on any timescale and for any inhomogeneity
of the system under consideration. The first term on the right-hand side is a
conventional flow term. The second nonlinear Vlasov term represents the
self-consistent field acting on the given (first) particle and defined by all
particles of the system. It is worth underlying an impossibility of getting
(without additional approximation) this term (showing up only for a space
inhomogeneous case) from the GME (or HGME) containing the time-independent
projection operator (see \cite{Los (2001)}). The term, quadratic in
$L_{12}^{^{\prime}}$, describes the collisions between particles and leads to
dissipation in the system. The terms with $L_{2}^{0}$ (derivative in space)
and with the time-derivative account for the change in space and time of the
one-particle distribution function on the microscopic (see also below) scales
of the orders of $r_{0}$ and $t_{cor}\sim r_{0}/\overline{v}$, respectively
($\overline{v}$ is the mean particle velocity). These two terms define the
contribution of the pair collisions to the nondissipative characteristics
(thermodynamic functions) of the non-ideal gas (see, e.g. \cite{Klimontovich
(1982)}) and are usually absent in the standard derivation of the kinetic
equations in the linear in $n$ approximation. These terms usually appear in
the next (second) approximation in $n$ (see, e.g. \cite{Uhlenbeck and Ford
(1963)}) that is not very consistent with the dissipative three-particle
collisions term showing up in this approximation. The last (irrelevant) term
in (\ref{27}) accounts for initial correlations (at $t=t_{0}$) and in the
linear approximation in $n$ is given by the two-particle correlation function
$g_{2}(x_{1},x_{2})$.

Let us look at the terms of (\ref{27}) more closely, especially at their
behavior in time. First of all, evolution with time is governed by the exact
two-particle propagator $G_{12}(t)=\exp[(L_{12}^{0}+L_{12}^{^{\prime}})t]$ (it
is natural for the considered dilute gas in the lowest approximation in
density), which satisfies the integral equation%
\begin{equation}
G_{12}(t)=G_{12}^{0}(t)+\int\limits_{0}^{t}dt_{1}G_{12}^{0}(t-t_{1}%
)L_{12}^{^{\prime}}G_{12}(t_{1}),\text{ } \label{28}%
\end{equation}
where $G_{12}^{0}(t)=\exp(L_{12}^{0})t$ is the propagator for non-interacting
particles. The action of $G_{12}^{0}(t)$ on any function defined on the phase
space is given by%
\begin{align}
e^{L^{0}t}\Phi(x_{1},\ldots,x_{N},t)  &  =\left[  \prod\limits_{i=1}%
^{N}e^{L_{i}^{0}t}\right]  \Phi(x_{1},\ldots,x_{N},t)\nonumber\\
&  =\Phi(\mathbf{x}_{1}-\mathbf{v}_{1}t,\mathbf{p}_{1},\ldots,\mathbf{x}%
_{N}-\mathbf{v}_{N}t,\mathbf{p}_{N},t), \label{29}%
\end{align}
which follows from (\ref{13}). Thus, if the particles' dynamics possesses the
necessary properties such as, e.g. a mixing ergodic flow in a phase space,
caused by the local (stochastic) instability, it may be expected that under
the action of $G_{12}(t)$ the distances between particles will rapidly
increase with time (a contribution of a "parallel motion" is negligible). If
the effective interaction between particles $V(\left\vert \mathbf{r}%
_{i}-\mathbf{r}_{j}\right\vert )$ vanishes at a distance $\left\vert
\mathbf{r}_{i}-\mathbf{r}_{j}\right\vert >r_{0}$, then, the integrand of the
integral over $t_{1}$ in (\ref{27}) vanishes at $t-t_{0}>t_{cor}$ and the same
is generally (but not always) valid for correlation function $g_{2}%
(x_{1},x_{2}) $ which also depends on the interaction between particles (in
more detail it has been considered in \cite{Los (2001)} for the space
homogeneous case).

On the other hand, the condition (\ref{24}) implies the existence of the time
hierarchy%
\begin{equation}
t_{cor}\ll t_{rel}, \label{30}%
\end{equation}
where $t_{rel}\thicksim\gamma^{-1}t_{cor}$ is the relaxation time for a
one-particle distribution function $F_{1}$($x_{i},t$) ($1/nr_{0}^{2}$ is a
particle's mean free path). In the initial stage of the evolution
$t_{0}\leqslant t\leqslant t_{cor}$, which is very interesting and essential
for understanding the irreversibility problem and for studying the
non-Markovian processes (the memory effects) and the ultrafast relaxation
effects, the initial correlations may be important. However, if one considers
the timescale $t-t_{0}\gg t_{cor}$, the initial correlations may damp and the
source in (\ref{27}) containing $g_{2}(x_{1},x_{2})$ may be disregarded. But
as Bogoliubov noted \cite{Bogoliubov (1946)}, this can be only done if the
time interval (\ref{0}) exists due to (\ref{30}). The latter conclusion
follows from the fact that the corrections to the solution of (\ref{27}), due
to initial correlations, are proportional to $\left\vert t-t_{0}\right\vert $
(secular terms) and are small only at $\left\vert t-t_{0}\right\vert \ll
t_{rel}$ (see \cite{Bogoliubov (1946)}). Thus, the expansion of the irrelevant
initial condition term in $n$ (as we did it while going from (\ref{22}) to
(\ref{27})) is ineffective because the terms of this expansion are
proportional to time and the convergence is very poor at the timescale of
interest in the kinetic theory $t-t_{0}\gtrsim t_{rel}$.

To get rid of the undesirable initial condition terms, it is often assumed
that $f_{i}(t_{0})=0$, which corresponds to the random phase approximation
(RPA) and is incorrect in principle (see \cite{van Kampen (2004)}).

Therefore (as it is already mentioned in the Introduction), in order to study
the kinetic stage of the evolution at $t\gtrsim t_{rel}$ a more sophisticated
approach like that based on the principle of weakening of initial correlations
(Bogoliubov's anzats) \cite{Bogoliubov (1946)} is needed. However, the
Bogoliubov approach does not allow for considering the initial stage of the
evolution $t_{0}\leqslant t\leqslant t_{cor}$ and the situations when the
large-scale correlations, associated, say, with the quantities that are
conserved, may exist. These correlations may only vanish on the timescale of
the order of the relaxation time $t_{rel}$. This approach also implies that
the correlations with $t_{cor}\gtrsim t_{rel}$ appearing in the higher
approximations in $n$ are not essential, although they are essential (see,
e.g. \cite{Ferziger and Kaper (1972), Klimontovich (1982)}).

If $F_{1}(x_{i},t)$ changes significantly on the macroscopic (hydrodynamic)
length $l_{h}$ and the corresponding timescale $t_{h}=l_{h}/\overline{v}$ (for
the space homogeneous gas there are only the characteristic scales $l$ and
$t_{rel}$ on which $F_{1}(x_{i},t)$ changes) the terms with $L_{2}^{0}$ and
$\frac{\partial F_{1}(x_{2},t-t^{^{\prime}})}{\partial t^{^{\prime}}}$ in
equation (\ref{27}) may be disregarded, because they are both at least of the
order of $\gamma$, i.e. their contribution to dissipative processes is of the
second order in $n$ (however, these terms should not be omitted if we wish to
account self-consistently for the nondissipative characteristics of the
nonideal gas). At the kinetic timescale, $t-t_{0}\gtrsim t_{rel}$, the upper
limit of integration over $t_{1}$ may be extended to infinity ($t_{0}%
\rightarrow-\infty$). If we also accept the RPA for $t_{0}\rightarrow-\infty$,
i.e. put $g_{2}(x_{1},x_{2})=0$, then equation (\ref{27}) reduces to%
\begin{align}
\frac{\partial F_{1}(x_{1},t)}{\partial t}  &  =L_{1}^{0}F_{1}(x_{1},t)+n\int
dx_{2}L_{12}^{^{\prime}}F_{1}(x_{2},t)F_{1}(x_{1},t)\nonumber\\
&  +n\int dx_{2}L_{12}^{^{\prime}}%
{\textstyle\int\limits_{0}^{\infty}}
dt_{1}e^{(L_{12}^{0}+L_{12}^{^{\prime}})t_{1}}L_{12}^{^{\prime}}F_{1}%
(x_{2},t)F_{1}(x_{1},t), \label{31}%
\end{align}
where we have taken in consideration that $F_{1}(x_{i},t-t_{1})$ may be
substituted (within the adopted approximation) with $F_{1}(x_{i},t)$ at
$t\gtrsim t_{rel}\gg t_{cor}$ (the integration over $t_{1}$ in (\ref{27})
gives essential contribution only up to $t_{1}\thicksim$ $t_{cor}$). In the
last (collision) term of (\ref{31}) the distribution functions of the collided
particles may be taken at the same space point $\mathbf{r}_{1}=\mathbf{r}_{2}$
with the adopted accuracy $\thicksim r_{0}/r_{h}\ll1$ (or $r_{0}/l\ll1$).

If the integral $%
{\textstyle\int\limits_{0}^{\infty}}
$ exists (and this is stipulated by the properties of the particles dynamics),
equation (\ref{31}) represents the nonlinear Markovian irreversible in time
kinetic equation. In the case of weak interaction between particles this
equation in the first approximation in the small interaction parameter
$\varepsilon$ gives the reversible Vlasov equation (equation (\ref{31}) with
disregarded the third collision term in the rhs, which is of the second order
in $\varepsilon$). In the second approximation in $\varepsilon$ it coincides
with the Vlasov-Landau kinetic equation (see, e.g. \cite{Balescu (1975)}). For
the space homogeneous case, when a one-particle distribution function does not
depend on a particle coordinate, $F_{1}(x_{j},t)=F_{1}(\mathbf{p}_{j},t)$
($\int F_{1}(\mathbf{p,}t)d\mathbf{p=}1$), the second (Vlasov) term in
(\ref{31}) vanishes, because we consider a potential $V_{ij}$ (\ref{13})
dependent on the particles' coordinates difference. In this case equation
(\ref{31}) is equivalent to the Boltzmann equation (see, e.g. \cite{Balescu
(1975)}). It should be underlined, that the nonlinearity of (\ref{31}) appears
quite naturally in this approach (based on time-dependent $P(t)$ (\ref{15}))
in contrast to the time-independent projection operator approach (see
\cite{Los (2001), Los (2005)}). In the latter case, we had to approximate
$F_{1}(x_{2},t_{0})$ by $F_{1}(x_{2},t)$ (it is possible on the timescale
given by (\ref{0})) in the evolution equation in order to get a nonlinear
equation from the linear GME or HGME.

\section{Homogeneous nonlinear generalized master equation retaining initial
correlations}

In section 2, the nonlinear generalized master equation for the relevant part
of a distribution function has been obtained. However, equation (\ref{12}) is
not closed in a sense that it is inhomogeneous one and contains an irrelevant
term comprising generally all multi-particle correlations. As was pointed
above, to get rid of this irrelevant part and obtain the closed equation for,
e.g., one-particle distribution function (like (\ref{31})), the Bogoliubov
principle of weakening of initial correlations or RPA are usually used.

In order to include initial correlations into consideration and to get the
evolution equations, valid on all timescales, the method turning the
conventional linear inhomogeneous generalized master equations (GMEs) into the
homogeneous form has been recently proposed in \cite{Los (2001), Los (2005)}.
This method has been applied to conventional linear GMEs (obtained by means of
the time-independent projection operator technique) and that resulted in the
linear homogeneous generalized master equations (HGMEs). Naturally, these
HGMEs are more convenient for studying the linear situations such as, e.g.
evolution of a small (sub)system interacting with a big system in an
equilibrium state (thermostat). For a many-particle system, such as considered
in the previous section, it is more convenient to deal with the obtained in
Sec.2 nonlinear generalized master equation (\ref{12}) (rather with its
simplified form (\ref{12l})). For example, as it was shown above, equation
(\ref{12l}) allows for obtaining the nonlinear equation for a one-particle
distribution function in the space inhomogeneous case which is not very easy
for consideration by means of the BBGKY hierarchy \cite{Bogoliubov (1946)}.

Now, we are going to apply the approach of \cite{Los (2001), Los (2005)} to
equation (\ref{12l}). Our goal is to obtain from (\ref{12l}) the homogeneous
equation containing the irrelevant part of a distribution function
(statistical) operator $f_{i}(t_{0})$ in the "mass" (super)operator acting on
the relevant part of a distribution function (statistical operator) $f_{r}%
(t)$. Such an equation will allow for treating the initial correlations
(contained in $f_{i}(t_{0})$) on the equal footing with all other correlations
(collisions) on any timescale including the initial stage of evolution when
initial correlations matter for sure. It will be the case, because the
perturbational expansion of a "mass" (super)operator does not lead to the
appearance of the "secular terms", proportional to time and violating the
perturbational expansion at the large (kinetic) timescale. It may be also
expected that the divergencies appearing in the higher order terms in the gas
density expansion \cite{Dorfman and Cohen (1967), Ferziger and Kaper (1972)}
will be more conveniently tackled by the "mass" (super)operator expansion. In
contrast to the HGMEs obtained in \cite{Los (2001), Los (2005)}, the equation
derived below is a nonlinear one and also applicable to the non-conservative
systems (with the time-dependent Hamiltonians).

Following \cite{Los (2001)}, let us identically present the irrelevant part of
a distribution function (statistical operator) as%
\begin{align}
f_{i}(t_{0})  &  =F(t_{0})-f_{r}(t_{0})=Q(t_{0})F(t_{0})\nonumber\\
&  =\left[  Q(t_{0})F(t_{0})\right]  F^{-1}(t_{0})U^{-1}(t,t_{0}%
)[P(t)+Q(t)]U(t,t_{0})F(t_{0})\nonumber\\
&  =C_{0}U^{-1}(t,t_{0})[f_{r}(t)+f_{i}(t)],\nonumber\\
C_{0}  &  =\left[  Q(t_{0})F(t_{0})\right]  F^{-1}(t_{0}),U^{-1}%
(t,t_{0})=T_{-}\exp[-%
{\textstyle\int\limits_{t_{0}}^{t}}
dsL(s)], \label{32}%
\end{align}
where $U^{-1}(t,t_{0})$ is the backward-in-time evolution operator for the
density matrix $F(t)$, $U^{-1}(t,t_{0})U(t,t_{0})=1$ (compare with (\ref{6})),
$T_{-}$ is the antichronological time-ordering operator arranging the
time-dependent operators $L(s)$ in such a way that time arguments increase
from left to right, $F^{-1}(t_{0})$ is invert to $F(t_{0})$, $F^{-1}%
(t_{0})F(t_{0})=1$, $P(t)+Q(t)=1$. Thus, the additional identity (\ref{32})
has been obtained by multiplying the irrelevant part by unity $F^{-1}%
(t_{0})F(t_{0})$ (which implies the existence of $F^{-1}(t_{0})$) and
inserting the unities $U^{-1}(t,t_{0})U(t,t_{0})=1$ and $P(t)+Q(t)=1$.
Therefore, neither divergency (caused by possible vanishing of $F(t_{0})$) nor
indetermination of the $0/0$ type (behaviors of the numerator and denominator
in $F(t_{0})/F(t_{0})=1$ are similar ) may happen. This holds over all further
(identical) manipulations (see below).

In (\ref{32}) the following parameter of initial correlations is introduced
\begin{align}
C_{0}  &  =[Q(t_{0})F(t_{0})]F^{-1}(t_{0})=f_{i}(t_{0})[f_{r}(t_{0}%
)+f_{i}(t_{0})]^{-1}\nonumber\\
&  =f_{i}(t_{0})f_{r}^{-1}(t_{0})[1+f_{i}(t_{0})f_{r}^{-1}(t_{0}%
)]^{-1}\nonumber\\
&  =(1-C_{0})f_{i}(t_{0})f_{r}^{-1}(t_{0}). \label{33}%
\end{align}
It is important to note that in (\ref{32}) and (\ref{33}) operator $Q(t_{0})$
acts only on $F(t_{0})$, which is reflected in comprising $Q(t_{0})F(t_{0})$
into the parenthesis. It follows from the fact that $f_{r}(t)$ and
$f_{i}(t_{0})$ are the basic quantities we are dealing with in equation
(\ref{12l}). All functions of dynamical variables, the average values of which
we can calculate with the help of $f_{r}(t)$ ($f_{red}(t)$) by multiplying
equation (\ref{12l}) with the corresponding functions (operators) from the
right and calculating an average value (a trace), are dependent only on the
variables which are not integrated off by $D$ ($D$ integrates off all
excessive variables in $F(t)$). Therefore, if we represent $f_{r}(t)$ and
$f_{i}(t_{0})$ in (\ref{12l}) as $f_{r}(t)=P(t)F(t)$ and $f_{i\text{ }}%
(t_{0})=Q(t_{0})F(t_{0})$, correspondingly, then the projection operators
$P(t)=C(t)D$ and $Q(t)=1-P(t)$ in these expressions act only on $F(t)$ but not
on the functions (if any) to the right of them. This is the essence of the
reduced description method, when, in order to calculate the average values of
the functions dependent on a much smaller number of variables than the whole
distribution function $F(t)$, we actually need only the reduced distribution
function (density matrix) $f_{red}(t)$.

As it is seen from (\ref{33}), the correlation parameter is a series in
$f_{i}(t_{0})f_{r}^{-1}(t_{0})$ and, therefore, one may only need a formal
existence of the function (operator) $f_{r}^{-1}(t_{0})$, which is invert to
the relevant distribution function (statistical operator) chosen with the help
of the appropriate operator $P(t_{0})$. It seems plausible that the invert
relevant part of a distribution function (statistical operator) defined in a
pointed above sense (uncorrelated part) may always be constructed (see
\cite{Los (2001), Los (2005)} and the next section).

Now, we have two equations, (\ref{10}) and (\ref{32}), relating $f_{i}(t)$ to
$f_{i}(t_{0})$. Using relations (\ref{12j}), finding $f_{i}(t_{0})$ from the
pointed equations as a function of $f_{r}(t)$ and inserting it into
(\ref{12l}), we obtain the following equation
\begin{align}
\frac{\partial Df_{r}(t)}{\partial t}  &  =DL^{0}(t)f_{r}(t)+DL^{^{\prime}%
}(t)R(t,t_{0})f_{r}(t)\nonumber\\
&  +DL^{^{\prime}}(t)R(t,t_{0})%
{\textstyle\int\limits_{t_{0}}^{t}}
dt^{^{\prime}}\overline{U}(t,t^{^{\prime}})\left[  Q(t^{^{\prime}%
})L(t^{^{\prime}})-\frac{\partial P(t^{^{\prime}})}{\partial t^{^{\prime}}%
}\right]  f_{r}(t^{^{\prime}}), \label{34}%
\end{align}
where the operator $R(t,t_{0})$ is defined as%
\begin{align}
R(t,t_{0})  &  =1+C(t,t_{0}),\nonumber\\
C(t,t_{0})  &  =\overline{U}(t,t_{0})\left[  1-C_{0}(t,t_{0})\right]
^{-1}C_{0}U^{-1}(t,t_{0}),\nonumber\\
C_{0}(t,t_{0})  &  =C_{0}U^{-1}(t,t_{0})\overline{U}(t,t_{0}), \label{35}%
\end{align}
$Df_{r}(t)=f_{red}(t)$ and relation (\ref{12h}) holds.

Equation (\ref{34}) is the central result of the paper. We have derived the
desirable homogeneous generalized evolution equation for the relevant part of
a distribution function (statistical operator). This equation differs from the
linear homogeneous time-convolution generalized master equation (TC-HGME)
obtained in \cite{Los (2001)} from the conventional Nakajima-Zwanzig TC-GME
with the help of time-independent projection operator technique. Equation
(\ref{34}) is generally a nonlinear one, valid for the systems with
time-dependent Hamiltonian and holds in both the classical and quantum physics
cases if the proper redefinition of the symbols is done and all
(super)operators exist (we will address the latter problem below). We have not
removed any information while deriving equation (\ref{34}), and, therefore, it
is exact integra-differential equation which accounts for initial correlations
and their dynamics through the modification of the (super)operator (memory
kernel) of (\ref{12l}) acting on the relevant part of a distribution function
(statistical operator) $f_{r}(t) $. The obtained exact kernel of (\ref{34})
may serve as a starting point for consecutive perturbation expansions. In many
cases such expansions of the homogeneous equations (like (\ref{34})) have much
broader range of validity than that of the inhomogeneous (like (\ref{12l}))
ones, when the expansions of the functions ($f_{r}$, $f_{i}$), rather than
equation, are involved (see also \cite{Bogoliubov (1946)}). Equation
(\ref{34}) reduces to the linear TC-HGME obtained in \cite{Los (2001)} if the
(projection) operator $P$ and the Hamiltonian $H$ do not depend on time and
relations (\ref{12h}), (\ref{12h'}) hold. Equation (\ref{34}) is suitable for
both, the many-body systems, evolution of which is described, naturally, by
the nonlinear (like the Boltzmann) equations, and the open systems like the
ones interacting with the thermostat (the linear version of (\ref{34})). How
the linear TC-HGME works in the classical and quantum physics cases has been
demonstrated in \cite{Los (2001)} and \cite{Los (2005)}.

The problem of the existence (convergency) of $R(t,t_{0})$ may be raised. The
function $R(t,t_{0})$ behaves properly at all times. Moreover, the expansion
of the kernel of (\ref{34}) may result in canceling the pole in the function
$R(t,t_{0})$. For linear TC-HGME it has been shown in \cite{Los (2001), Los
(2005)} in the linear approximation in the small density for a space
homogeneous dilute gas of classical and quantum particles. In such cases there
is no problem with the existence of $R(t,t_{0})$. Looking at (\ref{33}) (from
which the relation $(1-C_{0})^{-1}C_{0}=f_{i}(t_{0})/f_{r}(t_{0})$ follows),
one may expect that the same will be the case for $R(t,t_{0})$ (\ref{35}) (see below).

\section{Evolution equation accounting for initial correlations for a dilute
gas of particles}

Let us now apply equation (\ref{34}) to a system of $N$ ($N\gg1$) identical
interacting classical particles described by the Liouvillian (\ref{13}) and
considered in Sec. 3. In order to get an equation for a one-particle
distribution function $F_{1}(x_{i},t)$ ($i=1,\ldots,N$), we will use the same
operator $P(t)$\ (\ref{15}). Thus, taking into account (\ref{13a}), (\ref{19})
and (\ref{20}), we obtain from (\ref{34}) the following equation for a
one-particle distribution function%
\begin{align}
\frac{\partial F_{1}(x_{1},t)}{\partial t}  &  =L_{1}^{0}F_{1}(x_{1},t)+n\int
dx_{2}L_{12}^{^{\prime}}F_{1}(x_{2},t)F_{1}(x_{1},t)+DL^{^{\prime}}%
C(t,t_{0})f_{r}(t)\nonumber\\
&  +DL^{^{\prime}}%
{\textstyle\int\limits_{t_{0}}^{t}}
dt^{^{\prime}}R(t,t_{0})\overline{U}(t,t^{^{\prime}})\left[  (L^{0}-L_{1}%
^{0})+Q(t^{^{\prime}})L^{^{\prime}}-\frac{\partial P(t^{^{\prime}})}{\partial
t^{^{\prime}}}\right]  f_{r}(t^{^{\prime}}). \label{38}%
\end{align}

Again, let us consider equation (\ref{38}) in the linear approximation in the
density parameter (\ref{24}). Given that action of $P(t)$ or $D$ results in
the expressions proportional to at least the first power in the density $n $
and that all terms on the rhs of (\ref{38}) (except the first flow term) are
already proportional to $n$ (or $D$), we may in the first approximation in $n$
disregard all terms with $P(t)$ in (\ref{38}) including such terms in
$\overline{U}(t,t^{^{\prime}})$. Then, in this approximation $\overline
{U}(t,t^{^{\prime}})=U(t,t^{^{\prime}})$ and is given by (\ref{25}).
Therefore, $U^{-1}(t,t_{0})\overline{U}(t,t_{0})$ in (\ref{35}) may be
approximated by unity and $C_{0}(t,t_{0})$ by $C_{0}$. Thus, in the adopted
approximation the correlation parameter
\begin{equation}
C(t,t_{0})=U(t,t_{0})[f_{i}(t_{0})f_{r}^{-1}(t_{0})]U^{-1}(t,t_{0}) \label{40}%
\end{equation}
as it follows from (\ref{33}) and (\ref{35}).

To calculate the contribution of initial correlations to the evolution
equation (\ref{38}), given by (\ref{40}), the definitions for $f_{r}(t_{0})$
(\ref{18}) and $f_{i}(t_{0})$ (\ref{18b}) should be used.

Now, proceeding as in Sec. 3., i.e. using (\ref{13a}), (\ref{19}), (\ref{26})
and the fact (pointed above) that each additional integration over the phase
space leads to the additional power of $n$, equation (\ref{38}) may be
presented in the first approximation in $n$ as%
\begin{align}
\frac{\partial F_{1}(x_{1},t)}{\partial t}  &  =L_{1}^{0}F_{1}(x_{1},t)+n\int
dx_{2}L_{12}^{^{\prime}}[1+C_{12}(t-t_{0})]F_{1}(x_{2},t)F_{1}(x_{1}%
,t)\nonumber\\
&  +n\int dx_{2}L_{12}^{^{\prime}}[1+C_{12}(t-t_{0})]%
{\textstyle\int\limits_{0}^{t-t_{0}}}
dt_{1}e^{L_{12}t_{1}}[(L_{2}^{0}+L_{12}^{^{\prime}})F_{1}(x_{2},t-t_{1}%
)F_{1}(x_{1},t-t_{1})\nonumber\\
&  +\frac{\partial F_{1}(x_{2},t-t_{1})}{\partial t_{1}}F_{1}(x_{1},t-t_{1})],
\label{41}%
\end{align}

where%
\begin{equation}
C_{12}(t-t_{0})=e^{L_{12}(t-t_{0})}\frac{g_{2}(x_{1},x_{2})}{F_{1}(x_{1}%
)F_{1}(x_{2})}e^{-L_{12}(t-t_{0})} \label{42}%
\end{equation}
$L_{12}=L_{12}^{0}+L_{12}^{^{\prime}}$, $L_{12}^{0}=L_{1}^{0}+L_{2}^{0}$ and
$F_{1}(x_{i})=F_{1}(x_{i},t_{0})$.

Homogeneous equation (\ref{41}) is the central result of this section. It
exactly describes (in the linear approximation in $n$) the evolution of a
one-particle distribution function of a nonideal classical gas of particles
with arbitrary space inhomogeneity and at any timescale. This equation exactly
(in the adopted approximation in $n$) accounts for initial correlations and
treats them on the equal footing with the collisions between particles and
other processes. Initial correlations enter the memory kernel (mass operator)
governing the evolution of the one-particle (reduced) distribution function
and do not result in any undesirable inhomogeneous term in the evolution
equation (\ref{41}) (compare with (\ref{27})). Contribution of initial
correlations to the evolution process is given by the time-dependent
correlation parameter (\ref{42}) which describes the evolution in time of the
normalized two-particle correlation function $\frac{g_{2}(x_{1},x_{2})}%
{F_{1}(x_{1})F_{1}(x_{2})}$. The time-evolution of all terms in (\ref{41}) is
defined by an exact two-particle propagator $G_{12}(t)=\exp(L_{12}t)$. Again,
it is worth stressing that equation (\ref{41}) is a nonlinear one, i.e.
contains the products of the distribution functions of the interacting
particles at the same time whereas equation, obtained in \cite{Los (2001)}
from the linear TC-HGME in the same, linear in $n$ approximation, is in \ fact
a linear equation for $F_{1}(x_{1},t)$ containing the products of $F_{1}%
(x_{1},t)$ and $F_{1}(x_{2},t_{0})$. For converting the latter equation into a
nonlinear one a coarse-graining over the interval $t-t_{0}\ll t_{rel}$ was
needed (on this time interval there is no difference between $F_{1}%
(x_{2},t_{0})$ and $F_{1}(x_{2},t)$).

Let us consider the rhs terms of (\ref{41}) and their evolution in time in
more detail. The first term is a conventional flow term caused by the gas
inhomogeneity. The second term represents the Vlasov self-consistent field
modified by initial correlations. If $C_{12}(t-t_{0})$ vanishes (under the
action of $G_{12}(t-t_{0})$) with time at $t-t_{0}\gtrsim t_{cor}$, we get the
conventional Vlasov term. The third term in the rhs of (\ref{41}) describes
the collisions between particles and the influence of the change of the
one-particle distribution function on the microscopic scales in space and time
on the collisions. All these processes are modified by initial correlations
given by $C_{12}(t-t_{0})$. Thus, equation (\ref{41}) accounts for influence
of all pair collisions and correlations on the dissipative and nondissipative
characteristics of the nonideal space inhomogeneous gas of particles.

Noting that the term with $L_{2}^{0}$ is of the order of $r_{0}/r_{h}$ and the
term with the time-derivative of $F_{1}(x_{2},t)$ is $\thicksim t_{cor}%
/t_{rel}$ ($\thicksim\gamma$), these terms may be disregarded (if one is not
interested in the nondissipative corrections to the nonideal gas dynamics) in
the adopted (linear in $n$) approximation. Then, equation (\ref{41}) takes the
simpler form%
\begin{align}
\frac{\partial F_{1}(x_{1},t)}{\partial t}  &  =L_{1}^{0}F_{1}(x_{1},t)+n\int
dx_{2}L_{12}^{^{\prime}}[1+C_{12}(t-t_{0})]F_{1}(x_{2},t)F_{1}(x_{1}%
,t)\nonumber\\
&  +n\int dx_{2}L_{12}^{^{\prime}}[1+C_{12}(t-t_{0})]\nonumber\\
&  \times%
{\textstyle\int\limits_{0}^{t-t_{0}}}
dt_{1}e^{L_{12}t_{1}}L_{12}^{^{\prime}}F_{1}(x_{2},t-t_{1})F_{1}(x_{1}%
,t-t_{1}), \label{43}%
\end{align}
where in the last (collision) term the distribution functions of the collided
particles may be taken at the same space point $\mathbf{r}_{1}=\mathbf{r}_{2}$
with the adopted accuracy $\thicksim r_{0}/r_{h}\ll1$ (in the case of plasma,
it is not good to do that in the second term due to the long-range Coulomb interaction).

If we consider the kinetic timescale, $t-t_{0}\gtrsim t_{rel}\gg t_{cor}$, and
the two-particle correlation function $C_{12}(t-t_{0})$ (\ref{42}) vanishes at
this timescale, equation (\ref{43}) may be (for the reasons outlined in the
Sec. 3) rewritten in the form of irreversible Markovian kinetic equation%
\begin{align}
\frac{\partial F_{1}(x_{1},t)}{\partial t}  &  =L_{1}^{0}F_{1}(x_{1},t)+n\int
dx_{2}L_{12}^{^{\prime}}F_{1}(x_{2},t)F_{1}(x_{1},t)\nonumber\\
&  +n\int dx_{2}L_{12}^{^{\prime}}%
{\textstyle\int\limits_{0}^{\infty}}
dt_{1}e^{L_{12}t_{1}}L_{12}^{^{\prime}}F_{1}(x_{2},t)F_{1}(x_{1},t).
\label{44}%
\end{align}
For an inter-particle interaction with the small parameter $\varepsilon\ll1 $,
the two-particle propagator $\exp(L_{12}t)$ in the collision term of
(\ref{44}) may be substituted with a "free" propagator $\exp(L_{12}^{0}t)$
(see (\ref{28})) and equation (\ref{44}) reduces in the second approximation
in $\varepsilon$ to the Vlasov-Landau equation (see, e.g. \cite{Balescu
(1975)}). In the space-homogeneous case equation (\ref{44}) coincides with the
Boltzmann equation for the momentum distribution function $F_{1}%
(\mathbf{p}_{1},t)$.

Note, that all stages of the evolution of the system under consideration may
be followed (in the adopted approximation) with the help of equation
(\ref{41}) beginning from the initial reversible regime $t_{0}\leqslant
t\leqslant t_{cor}$. If the reversible terms in (\ref{41}), related to initial
correlations $C_{12}(t-t_{0})$, vanishes with time and the integral $%
{\textstyle\int\limits_{0}^{\infty}}
$ exists, then the evolution of the system may switch from reversible to
irreversible stage described by, e.g., kinetic equation (\ref{44}). Thus, we
have developed an approach to studying the system's evolution which includes
the evolution in time of initial correlations and does not rely on the RPA or
the principle of weakening of initial correlations.

\section{Conclusion}

In this paper, the nonlinear generalized master equations describing the
evolution of the relevant part of a distribution function (statistical
operator) have been derived. This new approach is based on the use of the
nonlinear time-dependent operator $P(t)$ defining the relevant part of a
distribution function (statistical operator). Generally, this operator is not
a projection operator. The inhomogeneous nonlinear generalized master equation
(\ref{12}) (and (\ref{12l})) may be viewed as a generalization of the
Nakajima-Zwanzig time-convolution generalized master equation to the case of
the time-dependent operators $P$ and $Q$. For the case of time-independent
projection operators $P$ and $Q$ this equation reduces to the conventional
TC-GME. The obtained inhomogeneous nonlinear GME is equally useful for
deriving both the nonlinear and linear evolution equations for the reduced
distribution function (statistical operator) of interest in contrast to the
linear TC-GME which is naturally more convenient for deriving the linear
evolution equations, e.g. for a subsystem interacting with a thermostat. Using
the obtained equation, we have derived the inhomogeneous nonlinear equation
(\ref{27}) for a one-particle distribution function in the linear
approximation in the small density parameter of a gas of classical particles.
This equation, valid for arbitrary space inhomogeneity, contains the space and
time changes of the one-particle distribution function on the microscopic
timescale (contributing to the nondissipative characteristics of a nonideal
gas) and an irrelevant part (a source) given by the initial two-particle
correlation function. Although the obtained inhomogeneous nonlinear GME allows
for deriving the closed nonlinear equation for a one particle distribution
function, the additional assumption like RPA or the Bogoliubov principle of
weakening of initial correlations is needed in order to get either the
Vlasov-Landau equation or the Boltzmann kinetic equation.

For including the initial correlations into consideration we have applied the
method suggested in \cite{Los (2001)} and \cite{Los (2005)} to converting the
inhomogeneous nonlinear generalized master equation into the homogeneous form.
The obtained exact homogeneous nonlinear generalized master equation
(\ref{34}) for the relevant part of a distribution function (statistical
operator) describes all stages of the (sub)system of interest evolution
including the initial stage when the initial correlations matter. For deriving
this equation, which may be used for getting both the nonlinear and linear
evolution equations for the reduced distribution functions, no approximation
like RPA or the principle of weakening of initial correlations has been used.
To test this equation, we have applied it to a space inhomogeneous dilute gas
of classical particles and obtained in the linear in the density approximation
a new homogeneous nonlinear evolution equation (\ref{41}) for a one-particle
distribution function retaining initial correlations in the memory kernel.
This equation differs from the one obtained (in the same approximation for the
space-homogeneous case) in \cite{Los (2001)}, which in fact is a linear
equation derived from the linear homogeneous generalized master equation.
Equation (\ref{41}) describes all stages of the evolution, accounts for all
two-particle correlations (collisions) and converts to the Boltzmann or
Vlasov-Landau equation on the appropriate timescale if all initial
correlations vanish on this timescale. It is expected, that the next terms in
the density parameter expansion (including possible divergencies) will be more
easily treated in the framework of the obtained homogeneous generalized master
equation than by conventional approaches \cite{Dorfman and Cohen (1967),
Ferziger and Kaper (1972)}.

\end{document}